\documentclass[12pt]{iopart}
\usepackage{graphicx}
\begin{document}
\title{Noise delayed decay of unstable states: theory versus numerical simulations}

%\author{N V Agudov$^{1,2,}$\footnote{agudov@rf.unn.ru}, R Mannella$^{3}$, A V Safonov$^{1,2,}$
%\footnote{safalex@newmail.ru}, B
%Spagnolo$^1$\footnote{spagnolo@unipa.it} }

%\author{N V Agudov\dag, R Mannella\ddag, A V Safonov\S, Spagnolo$\|$
%\footnote[4]{spagnolo@unipa.it} }

\author{N V Agudov\dag \ddag, R Mannella\S, A V Safonov\dag \ddag\ and B Spagnolo\dag
\footnote[4]{spagnolo@unipa.it} }

\address{\dag\ INFM - Unit\`a di Palermo, Group of Interdisciplinary Physics and Dipartimento di Fisica e Tecnologie Relative,
Universit\`a di Palermo, Viale delle Scienze, I-90128 Palermo,
Italia}
\address{\ddag\ Radiophysical Department, State University of
N.Novgorod, 23 Gagarin Ave., Nizhny Novgorod, 603600, Russia}
\address{\S Dipartimento di Fisica and INFM UdR Pisa, Universit\`a degli Studi di Pisa, 56100 Pisa, Italia}

%\eads{\mailto{1}, \mailto{2}}

%\address{$^1$INFM - Unit\`a di Palermo, Group of Interdisciplinary Physics and Dipartimento di Fisica e Tecnologie Relative,
%Universit\`a di Palermo, Viale delle Scienze, I-90128 Palermo,
%Italia\\ $^2$Radiophysical Department, State University of
%N.Novgorod, 23 Gagarin
%Ave., Nizhny Novgorod, 603600, Russia\\
%$^3$ Dipartimento di Fisica and INFM UdR Pisa, Universit\`a degli Studi di Pisa, 56100 Pisa, Italia}

\date{\today}
\begin{abstract}
We study the  noise delayed decay of unstable nonequilibrium states in nonlinear
dynamical systems within the framework of the overdamped Brownian motion model.
We give the exact expressions for the decay times of unstable states for polynomial
potential profiles and  obtain nonmonotonic behavior of the decay times as a function
of the noise intensity for the unstable nonequilibrium states. The analytical results
are compared with numerical simulations.
\end{abstract}
\pacs{05.40.-a,02.50.-r,05.10.Gg}
%\eads{\mailto{1}, \mailto{2}}
\maketitle
\section{Introduction}
In the last two decades a large variety of noise induced phenomena
have been discovered in far from equilibrium nonlinear systems.
Among them there are stochastic resonance \cite{Gam98}, resonant
activation \cite{Doe92}, noise enhanced stability
\cite{ManS96}-\cite{AguS01} and noise delayed decay
\cite{AguMa95}-\cite{AguMa99}. Recently it was predicted
theoretically that the presence of additive noise may increase the
average escape time from metastable states
\cite{AguS99,AguS01,AguDS03,FiVaS03} and from unstable states
\cite{Agu98,AguMa99}. Moreover the noise can enhance the stability
of an unstable fixed point in nonlinear discrete maps
\cite{AguMa95,Wac99}. These resonance-like phenomena show a
nonmonotonic behaviour of the average escape time as a function of
the noise intensity. This means that by varying the noise
intensity we can lengthen or shorten the lifetime of unstable
states. In previous studies in fact was found that fluctuations
can only accelerate the escape time from unstable states
\cite{Are82}-\cite{Col91}. In the present work we study the
phenomenon of Noise Delayed Decay (NDD) in more detail by
analytical and numerical methods. Besides the Mean First Passage
Time (MFPT), which quantifies only one of many possible
characteristics of the decay process from unstable states, we
consider the Nonlinear Relaxation Time (NLRT). The NLRT takes into
account the inverse probability current directed from the stable
state to the unstable one, which is neglected by the MFPT. The
analytical calculations show that NDD effect for the NLRT is much
greater than that for MFPT \cite{AguMa99}. In the present paper
these analytical predictions are verified with numerical
simulations. We show that the NLRT can be increased by the noise,
when the noise intensity is varied in a very wide range. We
consider the model of one-dimensional overdamped Brownian motion
in the potential field of force
\begin{equation}\label{lan}
{dx\over dt}=-{d\Phi(x)\over\eta dx}+\xi (t).
\label{Langevin}
\end{equation}
Here $x$ is the coordinate of the Brownian particle or the
representative phase point denoting the state  of the system,
$\Phi(x)$ is the potential describing the system itself, $\xi(t)$
is the white Gaussian noise, with $\langle \xi(t)\rangle=0$,
$\langle\xi(t)\xi(t+\tau)\rangle=2q\delta(\tau)/\eta$, $2q/\eta$
is the intensity of fluctuations, $\eta$ is the coefficient of
equivalent viscosity, and $q=kT$ is the energy temperature of
fluctuations. The simplest case where the NDD appears is the
parabolic potential. This potential profile was considered in
theoretical papers concerning the decay from equilibrium unstable
states \cite{Are82,Haa81},\cite{Col91}-\cite{Ciu93}
\begin{equation}
\Phi(x)=-ax^2/2.
\label{phipar}
\end{equation}
The initial state is an unstable equilibrium state, if $x_0=0$,
and it is an unstable nonequilibrium state if $x_0\not=0$. Various
authors considered the decay in this system starting only from one
point $x_0=0$. In this case, the deterministic decay time is
infinity and the action of the noise decreases the decay time in
accordance with known scaling laws \cite{Are82}-\cite{Suz76}. On
the other hand, the effect of the NDD always appears when we
consider any $x_0\not=0$. In this case the decay times (both MFPT
and NLRT) growth with noise, reach some maximum, and decrease.
This effect was predicted theoretically in
references~\cite{Agu98,AguMa99}. Before we consider the analytical
expressions for MFPT and NLRT let us describe in more detail their
definitions. These definitions involve the decision interval
restricted by one or two boundaries which specify the area of the
unstable state. The  MFPT is the average time that the Brownian
particle stays within the decision interval before it crosses any
of the absorbing boundaries for the first time. Due to the
absorbing boundary conditions, the MFPT neglects that the particle
may return into the decision interval after it has crossed the
boundary once. On the other hand, the NLRT is an average time
which takes into account this inverse flow directed inside the
decision interval, when the particle comes back. The definition of
MFPT is well known and it is a very useful time characteristic of
decay processes, because the analytical expression for MFPT can be
written in quadratures for an arbitrary potential profile (See
e.g. references~ \cite{Str63}-\cite{And33}).
 Let $P(t)$ be the probability that the particle is
within the decision interval. Initially $P(0)=1$. With time the particle escapes from
the unstable state, therefore $P(\infty)=0$. If during the decay
process the particle may cross the boundaries of the decision interval any number of
times, the definition of the NLRT reads
\begin{equation}
\tau=\int_0^\infty P(t) dt.
\label{def}
\end{equation}
If the particle can cross the boundaries of the decision interval
only once, the time (\ref{def}) coincides with the MFPT. Otherwise
the expression  (\ref{def}) takes into account the inverse
probability current across the  boundaries of the decision
interval and differs from the MFPT. Therefore the MFPT is a
particular case of the NLRT (\ref{def}), namely, the case in which
we neglect the inverse probability current. The analytical
expressions for NLRT in quadratures for an arbitrary potential
were obtained in \cite{Mal97}.

All numerical simulations shown in the paper were done using the
 algorithms described in \cite{Mann02}, using the Heun algorithm for
 the stochastic integration, and the Ziggurath and a carry and subtract algorithm
for the generation of the noise deviates. The integration time
step used and the number of averages taken depend on the actual
numerical experiment carried out,
 and on the set of parameters studied. Both have been changed throughout
 the numerical experiments aiming at optimizing the simulations: for
 example, when the noise intensity $q$ was changed, the integration time
 step $h$ was varied accordingly, to keep the quantity $q h$ small, which
 is a prerequisite to have good convergence in the Heun integrator.

\section{Decay times for the symmetric potential}
For the symmetric potential profile $\Phi(-x)=\Phi(x)$, the MFPT
equals to (See e.~g. \cite{And33,Gar85})
\begin{equation}
T(x_0,L;q)={\eta\over q}\int_{x_0}^L e^{\Phi(v)/q}\int_0^v e^{-\Phi(u)/q}dudv.
\label{MFPTsym}
\end{equation}
Let us consider the polynomial potential profile
\begin{equation}
\Phi(x)=-a x^k/k,
\label{pot}
\end{equation}
where $k=2n$ is even. The exact expression for MFPT
(\ref{MFPTsym}) with potential (\ref{pot}) was obtained in
\cite{Are82}) for a particular case, when the initial state is
unstable equilibrium one $x_0=0$
\begin{equation}
T(0,L;q)={\eta L^2\over2q} {_2F_2} \left( 1,{2\over k};1+{1\over
k},1+{2\over k};{\Phi(L)\over q} \right),
\label{arecchisym}
\end{equation}
where $_2F_2(a_1,a_2;b_1,b_2;z)$ is a generalized hypergeometric
function \cite{Gra80,Abram}. It follows from (\ref{MFPTsym}), that
 for an arbitrary $x_0$
($-L<x_0<L$) the MFPT reads
\begin{equation}
T(x_0,L;q)=T(0,L;q)-T(0,x_0;q),
\label{MFPTsym2}
\end{equation}
For example, if the potential is parabolic, then $k=2$ and MFPT is
\begin{eqnarray}
T(x_0,L;q)={\eta \over2q}\left[ L^2 {_2F_2} \left( 1,1;{3\over
2},2;{\Phi(L)\over q} \right)-
\right.\nonumber\\
\label{MFPTpar}
\left.
x_0^2 {_2F_2} \left( 1,1;{3\over
2},2;{\Phi(x_0)\over q} \right) \right].
\end{eqnarray}
When the noise is absent the MFPT coincides with the deterministic decay time $T_d$.
In particular, for the parabolic potential the deterministic decay time is
$$
T_d(x_0,L)=(\eta/a)\ln (L/x_0).
$$
\begin{figure}
\begin{center}
\includegraphics[height=5cm]{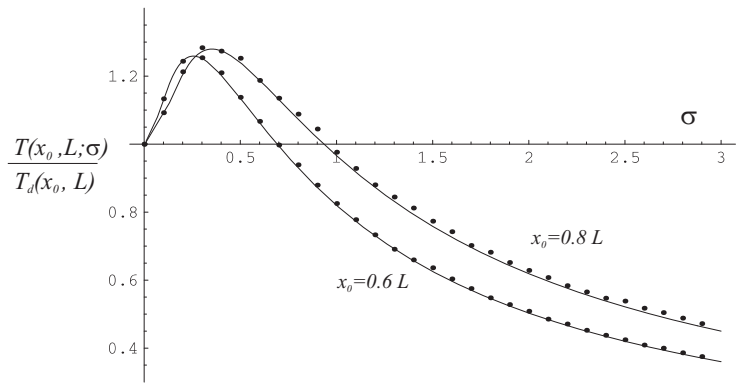}
\caption{The MFPT normalized to the deterministic decay time
versus the dimensionless temperature $\sigma=q/|\Phi(L)|$ for
non-equilibrium states described by parabolic potential, for two
initial states: $x_0=0.6L$ and $x_0=0.8L.$ Symbols are the results
of numerical simulations and solid lines are the theoretical
results obtained from equation (\ref{MFPTpar}).}
\end{center}
\end{figure}
The plots of the MFPTs normalized to $T_d$ for the parabolic
potential and for different $x_0$ are shown in figure 1, where the
symbols are the results of digital simulations and solid lines are
the theoretical predictions.
 One can see that the NDD effect appears. The agreement between theory
and numerical simulation is very good.
To take into account the influence of the inverse probability
current on the decay time
 we should consider NLRT. To obtain the NLRT we use the exact expression in
quadratures obtained in references~\cite{AguMa99,Mal97} for the
symmetric potential
\begin{equation}
\tau(x_0,L;q)=T(x_0,L;q)+\Theta(L;q),
\label{NLRT}
\end{equation}
where $T(x_0,q)$ is the MFPT (\ref{MFPTsym}) and
\begin{equation}
\Theta(L;q)={\eta\over q}
\int_L^\infty e^{\Phi(v)/q}dv
\int_0^L e^{-\Phi(v)/q}dv.
\label{Thetasym}
\end{equation}
For the polynomial potential (\ref{pot}) the NLRT correction
(\ref{Thetasym}) reads
\begin{eqnarray}
\Theta(L;q)={\eta L^2\over q}{(-1)^{-1/k}} \left(-\sigma
\right)^{2\over k} \label{Thetapoly} &\Gamma\left({1\over
k};\sigma^{-1}\right) \gamma\left({1\over k};-\sigma^{-1}\right),&
\end{eqnarray}
%\begin{eqnarray} &\Theta(L;q)={\eta L^2\over q}{(-1)^{-1/k}}
%\left({q\over\Phi(L)}\right)^{2\over k}\times &
%\nonumber\\
%\label{Thetapoly} &\Gamma\left({1\over k};{-\Phi(L)\over q}\right)
%\gamma\left({1\over k};{\Phi(L)\over q}\right),&
%\end{eqnarray}
%
%
where $\gamma(\alpha ;a)$ is the incomplete gamma function
\cite{Abram} and $\sigma =q/|\Phi(L)|.$ In particular, for the
parabolic potential profile (\ref{phipar}), when $k=2$, the NLRT
correction can be expressed in terms of error functions as
\begin{equation}
\Theta(L;q)={\eta\pi\over2a} \left(1-{\rm
Erf}(\sigma^{-1/2})\right) {\rm Erfi}(\sigma^{-1/2})
\label{Thetapar}
\end{equation}
The plots of NLRT are shown in
 figure 2, where symbols are the results of digital simulations and solid
lines are the theoretical predictions. The NLRT increases with
noise and displays the NDD effect. The enhancement of NLRT by
noise is much greater than that of MFPT. This is because of the
influence of the inverse probability current. Indeed, the NLRT
correction (\ref{Thetasym}) is  positive for any $q>0$.

\begin{figure}
\begin{center}
\includegraphics[height=5cm]{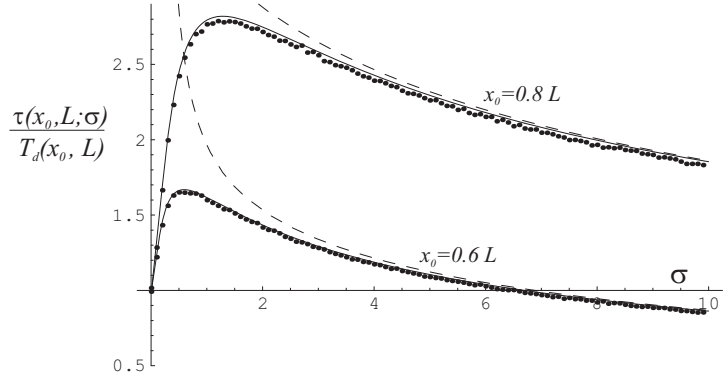}
\caption{The normalized NLRT versus the dimensionless temperature
$\sigma=q/|\Phi(L)|$ for non-equilibrium states described by a
parabolic potential, for two initial conditions, namely $x_0=0.6L$
and $x_0=0.8L.$ Symbols are the results of numerical simulations
and solid lines are the theoretical predictions.}
 \end{center}
\end{figure}
Note that the NLRT may exceed its deterministic value for a great
range of the noise temperature, e.~g. for $x_0=0.6L$ this range is
$0<q<6 |\Phi(L)|$ and for $x_0=0.8L$ this range is more than one
decade of dimensionless temperature. The value $|\Phi(L)|$ is the
maximal height of the potential profile within the decision
interval. Thus, the NDD effect takes place for the NLRT even for a
strong noise intensity, while for the MFPT it appears only for
weak noise when $q<|\Phi(L)|$. When $q>3|\Phi(L)|$ the
asymptotical expression may be obtained for NLRT
\begin{eqnarray}\nonumber
\tau(x_0,L;q)\approx {\eta\over aL^{k-2}} \left[
\Gamma\left({1\over k}\right)\sigma^{-1+1/k}- \right.
\\
\left. {k\over2}(1+m^2)\sigma^{-1} + {\Gamma\left({1\over
k}\right)\over k+1}\sigma^{-2+1/k}+ 0(\sigma^{-2}) \right],
\label{asymsym}
\end{eqnarray}
where $m=x_0/L$. The asymptotic (\ref{asymsym}), normalized to the
deterministic decay time $T_d(x_0,L)$ is shown in figure 2 by
dashed curve. It follows from equation~(\ref{asymsym}) that when
$\sigma\gg1$, the NLRT does not depend on the initial conditions,
i.e. on $x_0$. This is not the case for MFPT,
 which is
\begin{equation}
T(x_0,L;q)\approx {\eta k\over 2aL^{k-2}} (1-m^2)\sigma^{-1},
\end{equation}
for $\sigma\gg1$. We can not see this effect from figure 2 because
the NLRT is normalized to different deterministic decay time.
Therefore we may conclude that for strong noise intensity, the
inverse probability current removes differences in decay times
caused by initial conditions.
\begin{figure}
\begin{center}
\includegraphics[height=5cm]{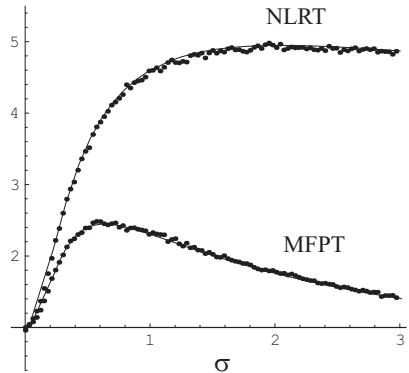}
\caption{The normalized MFPT and NLRT versus temperature
$\sigma=q/|\Phi(L)|$ for non-equilibrium state described by the
cubic potential $\Phi(x)=-x^3/3$, afor $x_0=0.8L.$ Symbols are the
results of numerical simulations and solid lines are the
theoretical predictions.}
\end{center}
\end{figure}
\section{Asymmetric potential}
Let us refer now to the asymmetric potential profile and consider
the polynomial potential (\ref{pot}) where $k=2n+1$ is odd. All
the states $x_0>0$ of this system are nonequilibrium. In reference
~\cite{Are82} the exact and approximate expressions for MFPT were
investigated for $x_0\leq0$ but not for $x_0>0$. While when
$x_0>0$ the effect of NDD appears. We consider the case $x_0>0$.
Let  the decision interval be $R:[-\infty;L]$, then  the MFPT
reads
\begin{equation}
T(x_0,L;q)={\eta\over q}\int_{x_0}^L e^{\Phi(v)/q}
\int_{-\infty}^v e^{-\Phi(u)/q}dudv.
\label{MFPTasim}
\end{equation}
and the NLRT, in accordance with \cite{Mal97}, is given by
equation (\ref{NLRT}) where
\begin{equation}
\Theta(L;q)=
{\eta\over q} \int_L^\infty e^{\Phi(v)/q}dv\cdot
\int_{-\infty}^L e^{-\Phi(u)/q}du.
\label{Thetaasym}
\end{equation}
Using the expression for $T(0,L;q)$ obtained in \cite{Are82} it is
possible to write the analytic expression for MFPT when $x_0>0$
\begin{eqnarray}
T(x_0,L;q)={\eta \over2q}
\left[
L^2
{_2F_2}\left(1,{2\over k};1+{1\over k},1+{2\over k};{\Phi(L)\over q}\right)
\right.&\nonumber\\
-x_0^2
{_2F_2}\left(1,{2\over k};1+{1\over k},1+{2\over k};{\Phi(x_0)\over q}\right)
\label{MFPTasym}\\
+
\left.
{k^{-2+2/k}\over q}\left({q\over a}\right)^{2\over k}
\Gamma\left({1\over k}\right)
\gamma\left({1\over k};{-\Phi(L)\over q},{-\Phi(x_0)\over q}\right)
\right]
.&\nonumber
\end{eqnarray}
The NLRT correction (\ref{Thetaasym}) for the polynomial potential (\ref{pot}) is
\begin{eqnarray}
\Theta(L;q)=(-1)^{1+1/k}\left(\eta \over q \right){\left(1\over
k\right)}^{2}{\left(qk\over a\right)}^{2/k}
\Gamma\left({1\over
k};-\sigma^{-1} \right) \Gamma\left({1\over k};\sigma^{-1}\right)
\label{Thetaasympoly}
\end{eqnarray}
The plots of MFPT (\ref{MFPTasym}) and NLRT (\ref{NLRT}) for a
cubic potential ($k=3$) and $x_0=0.8L$ are presented in figure 3,
where the symbols are the results of numerical simulation. We see
that the NDD takes place. The NLRT is always greater than the MFPT
because the inverse probability current always increase the decay
time. In general, despite the apparent difference between the
expressions for decay time in antisymmetric potentials, the
features of the times are similar.
\begin{figure}
\begin{center}
\includegraphics[height=5cm]{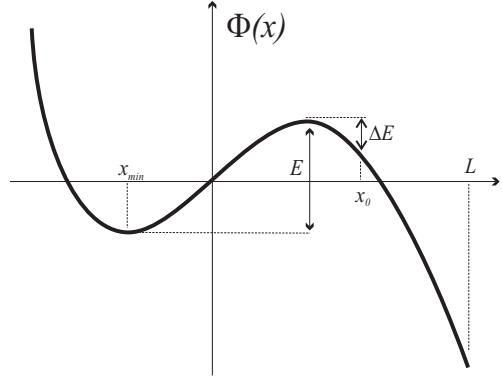}
\caption{Potential profile (\ref{barrier}).}
\end{center}
\end{figure}
\section{Potential with barrier}
Let us consider now a smooth potential profile with a barrier
\begin{eqnarray}
\Phi(x)=-\frac{a x^3}{3} + b x
\label{barrier}
\end{eqnarray}
where $a=1, b=1$ and boundary is located at the point $L=5$ (See
the figure 4). Then, the local minimum of this profile is located
at $x=-1,$ and the local maximum at $x=1.$ The barrier height is
$E=\Phi(1)-\Phi(-1)=4/3.$ The expressions in quadratures for MFPT
and NLRT for potential profile of this type were obtained in
\cite{AguMa99,Mal97}. They coincide with the corresponding
expressions for the antisymmetrical potential (\ref{NLRT}),
(\ref{MFPTasim}), and (\ref{Thetaasym}). We will consider two
cases of initial positions: $x_0=1.5$ and $x_0=3$. Both of these
initial states are situated behind the potential barrier. The
barrier height "seen" by the Brownian particle in its initial
position $x_0$ is $\Delta E=\Phi(1)-\Phi(x_0)$ (see figure 4).
Therefore $\Delta E/E<1$ for $x_0=1.5$ and $\Delta E/E>1$ for
$x_0=3$.  In references~\cite{AguS99,AguS01} it was shown that the
behaviours of MFPT and NLRT have strong dependence on the value
$\Delta E/E$. Indeed, if $\Delta E/E>1$, in the deterministic
limit (i.~e. when $q\to 0$) we obtain a finite value for the
escape time
\begin{equation}
T_d(x_0,L)= \eta\ln\left[\left(\frac{x_0 + 1}{L + 1}\right)
\left(\frac{L - 1}{x_0 - 1}\right)\right]^{1/2}.
\label{Td}
\end{equation}
By increasing the noise intensity the particles can go towards the
well and be delayed there. As a consequence, the MFPT and the NLRT
increase with $q$ and go through a maximum for some $q^*>0.$ This
is the typical case of NDD phenomenon considered above (see figure
5). However, if $\Delta E/E<1$ (see figure 6) the situation
becomes more complicated. It is evident that if there is no noise
($\xi=0$), the escape times (both MFPT and NLRT) are equal to the
deterministic escape time (\ref{Td}). At the same time, it follows
from equation~(\ref{MFPTasim}) that in the limit of $q\to 0$, the
MFPT and the NLRT go to infinity and for $q=0$ a singularity
appears. This singularity was demonstrated for the first time in
references~\cite{AguS99,AguS01}. The asymptotic expression for the
MFPT as $q\to0$ and $E>\Delta E>0$ can be obtained from
equation~(\ref{MFPTasim}):
$$
T(x_0; q)\approx \sqrt{\frac{2\pi q}
{\Phi''(x_{min})}}
\frac{\eta}{(-\Phi '(x_0))}
\exp\left (
\frac{E-\Delta E}{q}
\right),
$$
where $x_{min}$ is the coordinate of the local minimum.
\begin{figure}
\begin{center}
\includegraphics[height=5cm]{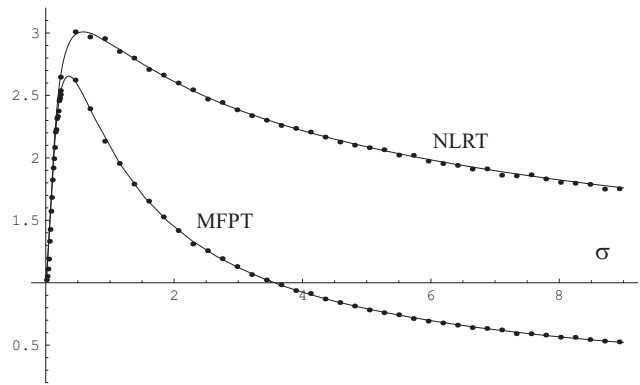}
\caption{The normalized MFPT and NLRT versus the dimensionless temperature
for potential profile with the
barrier (\ref{barrier}) and for non-equilibrium initial unstable state $x_0=3$ ($\Delta E>E$).
Symbols are the results of numerical
simulations and solid
lines are the theoretical predictions.}
\end{center}
\end{figure}
\begin{figure}
\begin{center}
\includegraphics[height=5cm]{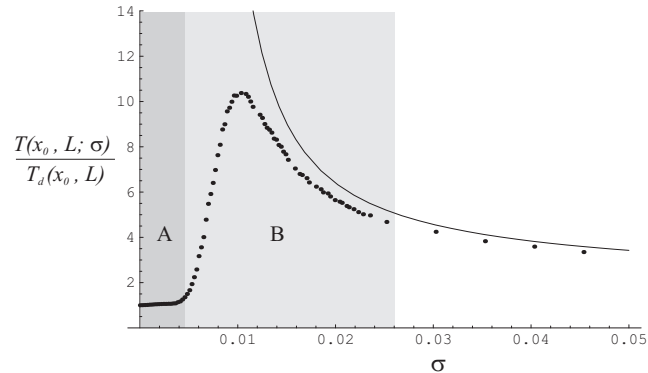}
\caption{The normalized MFPT versus the dimensionless
temperature $\sigma=q/E$
for potential profile with the
barrier (\ref{barrier}) and for non-equilibrium initial unstable state
$x_0=1.5$ ($\Delta E<E$).
Symbols are the results of
numerical simulations and solid
lines are the theoretical predictions.}
\end{center}
\end{figure}
For both above cases the plots of MFPT and NLRT are shown in
figures 5 and 6, where the symbols are the results of digital
simulations and solid lines are the theoretical predictions. One
can see, that the agreement between theory and the results of
numerical simulations is good only for initial position $x_0=3$
and there is strong difference when $x_0=1.5$. In the last case,
when $q \to 0$, the theoretical curve goes to infinity, while
digital simulation gives the deterministic time. We can interpret
this effect as follows: as explained in reference~\cite{AguS01}
the theoretical curve goes to infinity because even a very small
noise intensity can eventually push the particle, initially
located behind the barrier, back into the potential well. Then,
the particle will be trapped there for a long time, because the
well is very deep. These trajectories of the Brownian particles
lead to the singularity of the MFPT for $q\to0$. On the other
hand, the observation time in a digital simulation is finite and
for some value of $q$ it becomes smaller than the average escape
time of a particle trapped in the well. Therefore the simulated
escape time has the maximum at this point (See the region B in
figure 6). Besides, the ensemble of particles in a numerical
experiment is also finite. That is why for very small $q$, when
the probability for a particle to be trapped in the well decreases
exponentially to zero, we do not observe such particles in the
simulation. As a consequence, the average decay time in simulation
becomes equal to the deterministic one at $q\to0$ (See the region
A in the figure 6).

\section{Conclusions}
 We have derived the exact expressions for MFPT and NLRT in the
case of polynomial potential profiles and for arbitrary unstable
non-equilibrium initial positions. Expressions for MFPT obtained
earlier (See e.g. \cite{Are82}) are valid only for the particular
case when the initial position is the unstable equilibrium state
and the action of noise always decreases the decay times. We have
obtained nonmonotonic behaviour of the decay times as a function
of the noise intensity and have shown that the inverse probability
current can increase the NLRT with respect to the MFPT. Moreover
the range in which we obtain the NDD effect is larger for the NRLT
than for the MFPT.
 An important characteristic of the NDD effect is that we can
both accelerate or slow down the decay of unstable nonequilibrium
states by varying the intensity of fluctuations in a larger range
for NRLT than for MFPT.
 The numerical simulations are in good agreement with the theory
for most of potentials and initial conditions considered, except
the specific case of singularity in the MFPT and the NLRT for
$q\to0$. This singularity appears for the non-equilibrium states
in a potential profile with a barrier when $\Delta E<E$ (figure 6)
 and it is also responsible for the enhancement of the
stability of fluctuating metastable states
\cite{AguS01,AguDS03,FiVaS03}. In this case due to the limitation
in the particles number and observation time in simulation, we
cannot obtain the singularity (infinite decay time) numerically.
Therefore one should expect that the same difference found between
theoretical values of the average decay times and numerical
simulations for $\Delta E<E$ and $q\to0$ will appear in
experimental results when compared with theoretical ones.

\section{Acknowledgments}
This work has been supported in part by INTAS Grant 01-450, by
INFM and MIUR, and by RFBR (Project Nos.~02-02-17517,
1729.2003.2).

\end{document}